\documentclass[square]{ws-procs11x85}
\usepackage{graphicx,color}
\usepackage{amsmath,amssymb,wrapfig}

\newcommand{\homo}{\mathsf{H}} 

\definecolor{darkgrn}{rgb}{0, 0.75, 0}

\newcommand{\R}{{\mathbb R}}

\newcommand{\vx}{ \mathbf{x} }

\usepackage[colorlinks,linkcolor=black,bookmarksopen,
bookmarksnumbered,citecolor=black,urlcolor=black]{hyperref}

\begin{document}
\title{TOPOLOGICAL FEATURES IN CANCER GENE EXPRESSION DATA}

\author{S.~LOCKWOOD}
\address{School of Electrical Engineering and Computer Science, Washington State University,\\ 
Pullman, WA 99164, U.S.A.\\ 
E-mail: \href{mailto:svetlana.lockwood@email.wsu.edu}{svetlana.lockwood@email.wsu.edu}}

\author{B.~KRISHNAMOORTHY}
\address{Department of Mathematics, Washington State University,\\ 
Pullman, WA 99164, U.S.A.\\ 
E-mail: \href{mailto:bkrishna@math.wsu.edu}{bkrishna@math.wsu.edu}}

\begin{abstract}
We present a new method for exploring cancer gene expression data based on tools from algebraic topology. Our method selects a small relevant subset from tens of thousands of genes while {\em simultaneously} identifying nontrivial higher order topological features, i.e., holes, in the data. We first circumvent the problem of high dimensionality by {\em dualizing} the data, i.e., by studying genes as points in the sample space. Then we select a small subset of the genes as landmarks to construct topological structures that capture persistent, i.e., topologically significant, features of the data set in its first homology group. Furthermore, we demonstrate that many members of these loops have been implicated for cancer biogenesis in scientific literature. We illustrate our method on five different data sets belonging to brain, breast, leukemia, and ovarian cancers.
\end{abstract}


\bodymatter

\section{Introduction}\label{sl:intro}
During the past few decades, science has made great progress in understanding the biology of cancer \cite{p53review_yu2014small,cancer_advances_2014}. The latest technological tools allow assaying tens of thousands of genes simultaneously, providing large volumes of data to search for cancer biomarkers \cite{microarray_techn_2013,venter2001sequence}. Ideally, scientists would like to extract some qualitative signal from this data in the hope to better understand the underlying biological processes. At the same time it is desirable that the extracted signal is robust in the presence of noise and errors while effectively describing the dataset \cite{valarmathi2014noise,pozhitkov2014revised}.

One suite of methods which have enjoyed increased level of success in recent years is based on concepts from the mathematical field of algebraic topology, in particular, {\em persistent homology} \cite{EdLeZo2002,Gh2008barcodes,CaZoCoGu2004}. The key benefits of using topological features to describe data include coordinate-free description of shape, robustness in the presence of noise and invariance under many transformations, as well as highly compressed representations of structures \cite{Lumetal2013}. Analysis of the homology of data allows detection of high-dimensional features based on {\em connectivity}, such as loops and voids, which could not be detected using traditional methods such as clustering \cite{Ca2009,Gh2008barcodes}. Further, identifying the most {\em persistent} of such features promises to pick up the significant shapes while ignoring noise \cite{Gh2008barcodes}. Such analysis of the stable topological features of the data could provide helpful insights as demonstrated by several studies, including some recent ones on cancer data \cite{dewoshkin2010,seeman2012}.

\subsection{Our contributions} \label{contrib}
We present a new method of topological analysis of various cancer gene expression data sets. Our method belongs to the category of exploratory data analysis. In order to more efficiently handle the huge number of genes whose expressions are recorded in such data sets (typically in the order of tens of thousands), we transpose the data and analyze it in its {\em dual space}, i.e., with each gene represented in the much lower dimensional (in the order of a few hundred)  sample space. We then sample critical genes as guided by the topological analysis. In particular, we choose a small subset (typically $120$--$200$) of genes as {\em landmarks} \cite{desilva2004}, and construct a family of nested simplicial complexes, indexed by a proximity parameter. We observe topological features (loops) in the first homology group ($\homo_1$) that remain significant over a large range of values of the proximity parameter (we consider small loops as topological noise). By repeating the procedure for different numbers of landmarks, we select stable features that persist over large ranges of both the number of landmarks and the proximity parameter. We then further analyze these loops with respect to their membership, working under the hypothesis that their topological connectivity could reveal functional connectivity. Through the search of scientific literature, we establish that many loop members have been implicated in cancer biogenesis. We applied our methodology to five different data sets from a variety of cancers (brain, breast, ovarian, and acute myeloid leukemia (AML)), and observed that in each of the five cases, many members of the significant loops in $\homo_1$ have been identified in the literature as having connections to cancer.

Our method is capable of identifying geometric properties of the data that cannot be found by traditional algorithms such as clustering \cite{cluster_genes_review2012,cluster_review2010}. By employing tools from algebraic topology, our method goes beyond clustering and detects connected components around holes (loops)  in the data space. The shown methodology is also different from techniques such as graph \cite{cluster_massive_graphs_siam2013,graph_mining_book_cook2006} or manifold learning \cite{Isomap_original2000,LLE_original2000,HessianLLE_original2003,laplacian_embedding_original2001,diffusion_maps2006}. Graph algorithms, while identifying connectivity, miss wealth of information beyond clustering. Manifold learning algorithms assume that the data comes from an intrinsically low-dimensional space and their goal is to find a low-dimensional embedding. We do not make such assumptions about the data.

\subsection{Related work}\label{prev_work}
Several applications of tools from algebraic topology to analyze complex biological data from the domain of cancer research have been reported recently. DeWoshkin et al.~\cite{dewoshkin2010} used computational homology for analysis of comparative genomic hybridization (CGH) arrays of breast cancer patients. They analyzed DNA copy numbers by looking at the characteristics of \(\homo_0\) group. Using $Betti_0$ (\(\beta_0\)) numbers, which are the ranks of the zeroth homology groups ($\homo_0$), their method was able to distinguish between recurrent and non-recurrent patient groups.

Likewise, Seeman et al.~\cite{seeman2012} applied persistent homology tools to analyze cancer data. Their algorithm starts with a set of genes that are preselected using the {\it nondimensionalized standard deviation} metric \cite{seeman2012}. Then, by applying persistent homology analysis to the \(\homo_0\) group, the patient set is recursively subdivided to yield three subgroups with distinct cancer types. By inspecting the cluster membership, a core subset of genes is selected which allows sharper differentiation between the cancer subtypes.

Another example of topological data analysis is the work of Nicolau et al.~\cite{nicolau2011}. Their method termed {\em progression analysis of disease} (PAD) is applied to differentiate three subgroups of breast cancer patients. PAD is a combination of two algorithms -- disease-specific genome analysis (DSGA) \cite{dsga2007} and the topology-based algorithm termed {\it Mapper} \cite{mapper2007}.  First, DSGA transforms the data by decomposing it into two components -- the disease component and the healthy state component, where the disease component is a vector of residuals from a {\it Healthy State Model} (HSM) linear fit. A small subset of genes that show a significant deviation from the healthy state are retained and passed on to Mapper, which applies a specified filter function to reveal the topology of the data. Mapper identified three clusters corresponding to ER$+$, ER$-$, and normal-like subgroups of breast cancer. This work is somewhat different from the previous two papers mentioned above because it does not explicitly analyze features of any of the homology groups.

All studies mentioned above utilized $\beta_0$ numbers, thus performing analyses that are topologically equivalent to clustering. In contrast, our method relies on $\beta_1$ numbers (ranks of $\homo_1$ groups). One can think of $\beta_1$ numbers characterizing the loops constructed from connected components (genes) around ``holes'' in the data. The underlying idea is that connections around holes  may imply connections between the participating genes and biological functions. Also, most of other methods use some data preprocessing to limit the initial pool of candidate genes. Our method selects the optimal number of genes as part of the analysis itself.

\section{Mathematical background} \label{sl:theory}

We review some basic definitions from algebraic topology used in our work. For details, refer one of the standard textbooks \cite{Munkres1984,EdHa2009}. Illustrations of simplices, persistent homology, and identification of topological features from landmarks are available in the literature \cite{Gh2008barcodes, desilva2004, javaplex2011}.

\subsection{Simplices and simplicial complexes} \label{sl:simplicial_complexes}
Topology represents the shape of point sets using combinatorial objects called simplicial complexes. Consider a finite set of points in \(\R^n\). More generally, the space need not be Euclidean. We just need a unique pairwise distance be defined for every pair of points. The building blocks of the combinatorial objects are {\em simplices}, which are made of collections of these points. 

Formally, the convex hull of $k+1$ affinely independent points $\{v_{0}, v_{1}, \dots, v_{k}\}$ is a $k$-simplex. The dimension of the simplex is $k$, and $v_{j}$s are its vertices. Thus, a vertex is a $0$-simplex, a line segment connecting two vertices is a $1$-simplex, a triangle is a $2$-simplex, and so on. Observe that each $p$-simplex $\sigma$ is made of lower dimensional simplices, i.e., $k$-simplices $\tau$ with \(k \leq p\). Here, $\tau$ is called a {\em face} of $\sigma$, denoted $\tau \subset \sigma$.  A collection of simplices satisfying two regularity properties forms a {\em simplicial complex}. The first property is that each face of every simplex in a simplicial complex $K$ is also in $K$. Second, each pair of simplices in $K$ intersect in a face of both, or not at all. Due to these properties, algorithms to study shape and topology run much more efficiently on the simplicial complex than on the original point set.

To construct a simplicial complex on a given point set, one typically considers balls of a given diameter $\epsilon$ (called $\epsilon$-ball) centered at each point. The two widely studied complexes of this form are the {\em \v{C}ech} and the {\em Vietoris-Rips} complexes. A $k$-simplex is included in the \v{C}ech complex if there exists an $\epsilon$-ball containing all its $k+1$ vertices. Such a simplex is included in the the Vietoris-Rips complex \(R_{\epsilon} \) if each pair of its vertices is within a distance $\epsilon$. As such, Vietoris-Rips complexes are somewhat easier to construct, since we only need to inspect pairwise, and not higher order, distances.

However, both the \v{C}ech and the Vietoris-Rips complexes have as vertex set all of the points in the data. Such complexes are computationally intensive for datasets of tens of thousands of points. The feasible option is to work with an approximation of the topological space of interest \cite{desilva2004}. The key idea is to select only a small subset of points (landmarks), while the rest of points serve as {\it witnesses} to the existence of simplices. Termed {\em witness complexes}, such complexes have a number of advantages. They are easily computed,  adaptable to arbitrary metrics, and do not suffer from the curse of dimensionality. They also provide a less noisy picture of the topological space. We use the {\it lazy Witness complex}, in which conditions for inclusion are checked only for pairs and not for higher order groups of points \cite{desilva2004}, analogous to the distinction between the constructions of Vietoris-Rips and \v{C}ech complexes.

We employ the heuristic landmark selection procedure called {\it sequential maxmin} to select a representative set of landmark points \cite{desilva2004,adams_carlsson_2009,carlsson_etal_2008}. The first landmark is selected randomly from the point set $S$. Then the algorithm proceeds inductively. If \(L_{i-1}\) is the set of the first \(i-1\) landmarks, then the $i$-th landmark is the point of $S$ which maximizes the function \(d(x, L_{i-1})\), the distance between the point \(x\) and the set \(L_{i-1}\). We vary the total number of landmarks, exploring each of the resulting lazy witness complexes. The final number of landmarks is chosen so that the resulting witness complex maximally exposes topological features.

\subsection{Persistent homology}\label{persist_homo}

Homology is the concept from algebraic topology which captures how space is {\em connected}. Thus, homology can be used to characterize interesting features of a simplicial complex such as connected clusters, holes, enclosed voids, etc., which could reveal underlying relationships and behavior of the data set. Homology of a space can be described by its {\em Betti numbers}. The $k$-th Betti number \(\beta_k\) of a simplicial complex is the rank of its $k$-th homology group. For $k=0,1,2$, the $\beta_k$ have intuitive interpretation. \(\beta_{0}\) represents a number of connected components, \(\beta_{1}\) the number of holes, and \(\beta_{2}\) the number of enclosed voids. For example, a sphere has $\beta_0=1, \beta_1=0, \beta_2=1$, as it has one component, no holes, and one enclosed void.

Consider the formation of a simplicial complex using balls of diameter $\epsilon$ centered on points in a set. For small \(\epsilon\), the simplicial complex is just a set of disjoint vertices. For sufficiently large \(\epsilon\), the simplicial complex becomes one big cluster. What value of \(\epsilon\) reveals the ``correct'' structure? Persistent homology \cite{EdLeZo2002,Gh2008barcodes} gives a rigorous response to this question. By increasing \(\epsilon\), a sequence of nested simplicial complexes called a filtration is created, which is examined for attributes of connectivity and their robustness. Topological features appear and disappear as \(\epsilon\) increases. The features which exist over a longer range of the parameter $\epsilon$ are considered as signal, and short-lived features as noise \cite{EdLeZo2002,zomo_carlsson_2005}. This formulation allows a visualization as a collection of barcodes (one in each dimension), with each feature represented by a bar. The longer the life span of a feature, the longer its bar. In the example barcodes in Figs.~\ref{breast_evolve}--\ref{aml170_dim25}, the x-axis represents the \(\epsilon\) parameter, and the bars of persistent loops of interest are circled.

\section*{Research questions}

Our approach could address several critical questions in the context of cancer data analysis. First, could we select a small subset of relevant genes while {\em simultaneously} identifying robust nontrivial structure, i.e., topology, of the data? Most previous approaches require the selection of a subset of genes {\em before} exploring the resulting structure, and hence limiting the generality. Second, could we elucidate higher order interactions (than clusters) between genes that could have potential implications for cancer biogenesis? Higher order structures such as loops could reveal critical subsets of genes with relevant nontrivial interactions, which together have implications to the cancer. Third, could this method work even when data is available from only a {\em subset} of patients?

\section{Data}\label{sl:data}
We analyzed five publicly available microarray datasets of gene expression from four different types of cancer -- breast, ovarian, brain, and acute myeloid leukemia (AML). Four of the datasets have the same protocol, GPL570 (HG\_U133\_Plus\_2, Affymetrix Human Genome U133 Plus 2.0 Array). The fifth dataset has a different protocol, HG\_U95Av2, which has a fewer number of genes (see Table~\ref{sl:tbl_data}). By including data sets from different protocols, we could verify that the topological features identified are not just artifacts of a particular protocol. 

\begin{wraptable}{r}{3.8in}
\tbl{Datasets used in the study.}
{
\begin{tabular}{@{}lllcc@{}}
\toprule
Dataset & Series & Protocol & \# Genes & \# Samples \\ \colrule
Brain & GSE36245 & GPL570 & 46201 & 46\\
Breast & GSE3744 & GPL570 & 54613 & 47\\ 
Ovarian & GSE51373 & GPL570 & 54613 & 28\\
AML188 & GSE10358 & GPL570 & 54613 & 188\\
AML170 & willm-00119 & HG\_U95Av2 & 12558 & 170\\ \botrule
\end{tabular}
} \label{sl:tbl_data}
\end{wraptable}
The number of genes represents the number of unique gene id tags defined by a protocol excluding controls. While the brain dataset is of the same protocol as the breast and ovarian datasets, the former one has fewer genes -- 46201 vs.~54613. This variability, however, did not affect our procedure to find topological features.

All datasets, except for AML170, were obtained from NCBI Gene Expression Omnibus \url{http://www.ncbi.nlm.nih.gov/geo/} in October 2013. AML170 was retrieved at the same time from the National Cancer Institute caArray Database at \url{https://array.nci.nih.gov/caarray/project/willm-00119}. 

\section{Methods}\label{sl:methods}
We work with the raw gene expression values. In particular, we do not log-transform them.
\subsection{Dual space of data}
Traditionally, gene expression data is viewed in its gene space, i.e., the expression profile of patient $i$ with $m$ genes is a point in $\R^m$, $\vx_i= [ x_{i_1}~x_{i_2}~\cdots~x_{i_m} ]$. Each \(x_{i_j}\) is an expression of gene $j$ in the patient $i$. For example, each patient in the Brain dataset is a point in $\R^{46201}$.

We analyze expression data in its dual space, i.e., in its {\it sample space}. Hence a gene $j$ is represented as a point in $\R^n$, $\vx_j = [ x_{j_1}~x_{j_2}~\cdots~x_{j_n} ]$, where $n$ is the number of samples or patients. Each \(x_{j_i}\) is the expression of the gene $j$ in the patient $i$. For example, in the same Brain dataset, the points now sit in $\R^{46}$ space. Hence we study gene expression across the span of all patients.

The key motivation for this approach is to handle the high dimensionality in a meaningful way for analyzing the {\em shape} of data. Given the small size of patients, one could efficiently construct a Vietoris-Rips complex using the set of pairwise distances between patients in the gene space (no need to choose landmarks). But such distances become less discriminatory when the number of genes is large \cite{meaningiful_nneighbor_beyer1999,concentration_ledoux2005}. Working in the dual space, we let our topological method select a manageable number of genes as landmarks to construct the witness complexes, which potentially capture interesting topology of the data. Hence we do not preselect a small number of genes before the topological analysis, as is done by some previous studies \cite{nicolau2011,seeman2012}.

\subsection{Choosing the number of landmarks}
For construction of the witness complexes, the number of landmarks has to be defined {\it a priori} (Sec.~\ref{sl:simplicial_complexes}). Hence the question becomes how many landmarks do we select? We let the data itself guide the selection of genes used as landmarks. If there is a significant loop 
\begin{wraptable}{r}{2in}
\vspace*{-0.1in}
\tbl{Numbers of landmarks selected in each dataset.}
{\begin{tabular}{@{}lc@{}}\toprule
Dataset & \# landmarks \\ \colrule
Brain & 120\\
Breast & 110\\ 
Ovarian & 200\\
AML188 & 150\\
AML170 & 130\\ \botrule
\end{tabular}}\label{sl:tbl_num_landmarks}
\end{wraptable}
feature in the data, it would persist through a range of landmarks in $\homo_1$ of the complexes. We reconstruct the topological space incrementing the number of landmarks
while observing appearance and disappearance of the topological features. Initially, there would be very few small, noisy features because of insufficient number of points. As the number of landmarks increases, some features stabilize, i.e., do not change much either in size or membership. Then they reach their maximal size, and start to diminish once some critical number of landmarks is exceeded (when the ``holes'' are all filled in). The ``optimal'' number of landmarks is chosen when the length of the bar representing the topological feature is maximal.

\begin{figure}[ht!]
\centering
\includegraphics[width=\textwidth]{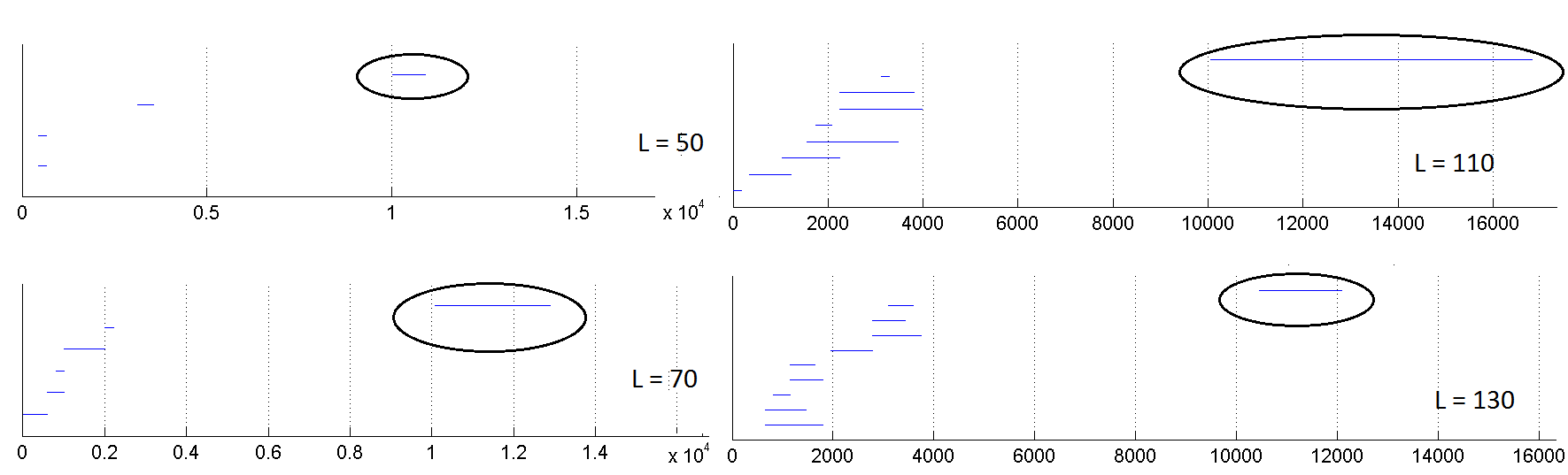}
\caption{Evolution of the loop of interest (circled) for varying number of landmarks from L=50 to L=130 in the breast dataset. Here and in Figs.~\ref{brain_loop}--\ref{aml170_dim25}, the x-axis represents the \(\epsilon\) parameter.}\label{breast_evolve}
\end{figure}

A typical example of such behavior is seen in the Breast dataset (see Fig.~\ref{breast_evolve}). A small loop appears when the number of landmarks is \(L=50\). It stabilizes around \(L=90\), reaches its maximum span at \(L=110\), and then decreases as \(L\) grows.

\subsection{Composition of loops}

One of the goals of our method is to determine the genes which participate in $\homo_1$ features, which could indicate potential implications for cancer biogenesis. Since the first landmark is chosen randomly in the sequential maxmin procedure, the composition of the loops identified may differ based on this first choice. To circumvent this effect, we do $20$ different runs in each case to collect possible variations in loop formation. Members of the loops are then pooled together for further analysis. Due to the almost deterministic nature of sequential maxmin selection (apart from the first landmark being selected randomly), we observed very little variation over the $20$ runs in most cases. The recovered members of loops are then queried in scientific literature for cancer-related reports.

\medskip
\noindent We implemented our computations using the package JavaPlex \cite{javaplex2011}. We explored the barcodes for $\homo_0, \homo_1,$ and $\homo_2$, but interesting persistent features with members related to cancer biogenesis were detected only for $\homo_1$.

\section{Results}\label{sl:results}

\begin{wrapfigure}{r}{3in}
\centering
\includegraphics[height=1.1in]{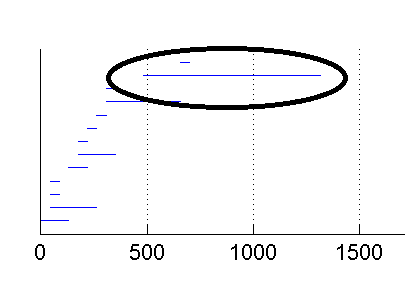}
\caption{Representative loop in brain dataset.}\label{brain_loop}
\end{wrapfigure}

\begin{wrapfigure}{r}{3in}
\centering
\includegraphics[width=3in]{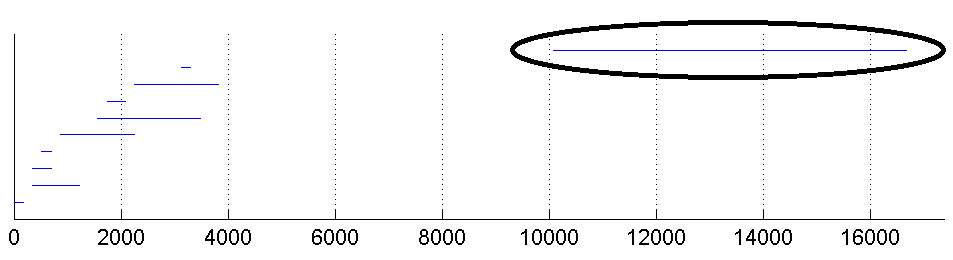}
\caption{Representative loop in breast dataset.}\label{breast_loop}
\end{wrapfigure}

\begin{wrapfigure}{r}{3in}
\centering
\includegraphics[width=3in]{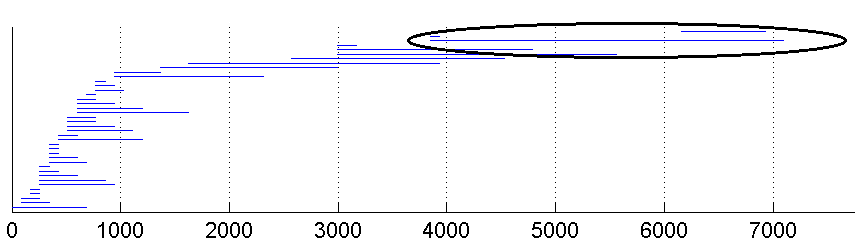}
\caption{Representative loop in ovarian dataset.}\label{ovarian_loop}
\end{wrapfigure}

\begin{wrapfigure}{r}{3in}
\centering
\vspace*{-0.1in}
\includegraphics[width=3in]{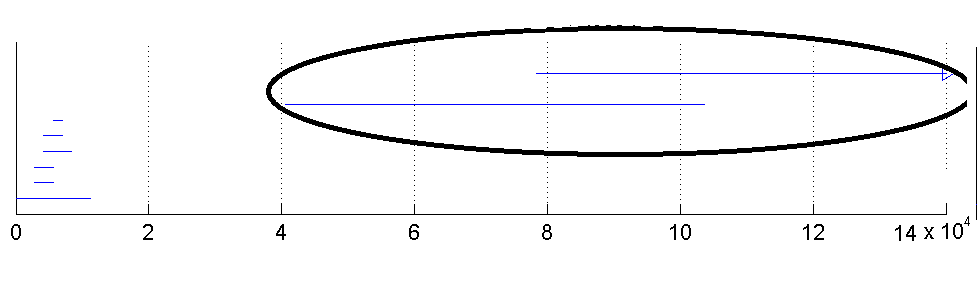}
\caption{Representative loops in AML188 dataset.}\label{aml188_2loops}
\end{wrapfigure}

\begin{wrapfigure}{r}{3in}
\centering
\vspace*{-0.1in}
\includegraphics[width=3in]{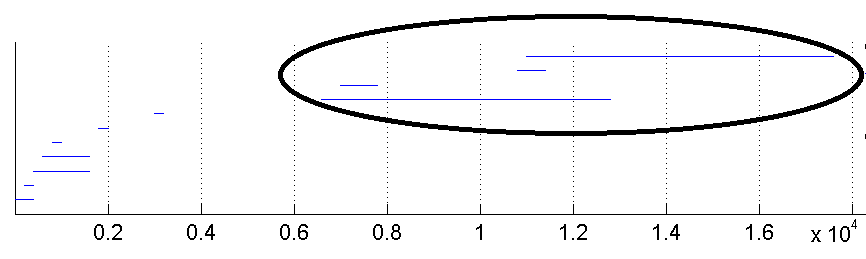}
\caption{Representative loops in AML170 dataset.}\label{aml170_2loops}
\end{wrapfigure}

Persistent topological features in the homology group \(\homo_1\) were observed in every cancer dataset we analyzed. Representative examples are shown in Figs.~\ref{brain_loop}--\ref{aml170_2loops}. The AML datasets both had two persistent loops, while the other datasets have one loop each. The Ovarian dataset had a few medium length bars in the $\homo_1$ barcode, but we investigated only the longest loop. Once the persistent loops were identified, they were inspected with respect to their composition and relation to cancer through search of scientific literature. Below is a brief description of results for each of the datasets (the full list of all loop members is also available \cite{suppl}).

\begin{table}[h]
\tbl{Selected representatives of loops in different datasets.}
{\begin{tabular}{@{}lllc@{}}\toprule
Gene & Dataset & Description & References \\ \colrule
CAV1 & Brain & tumor suppressor gene & \cite{gene_cav1}\\
RPL36 & Brain & prognostic marker in hepatocellular carcinoma & \cite{rpl36_2011}\\ 
RPS11 & Breast & downregulation in breast carcinoma cells & \cite{rps11_2001}\\
FTL & Breast & prognostic biomarkers in breast cancer & \cite{ftl_2006}\\
LDHA & Ovarian & overexpressed in tumors, important for cell growth & \cite{ldha_2013}\\
GNAS & Ovarian & biomarker for survival in ovarian cancer & \cite{gnas_2010}\\
LAMP1 & AML170 & regulation of melanoma metastasis & \cite{lamp1_2014}\\
PABPC1 & AML170 & correlation with tumor progression & \cite{pabpc1_2006}\\
HLF & AML188 & promotes resistance to cell death & \cite{hlf_2013}\\
DTNA & AML188 & induces apoptosis in leukemia cells & \cite{dtna_2012}\\ \botrule
\end{tabular}}\label{sl:tb_some_members}
\end{table}

\vspace*{-0.2in}
\subsection{Brain dataset}\label{sl:brain}
The lifespan of the longest loop in this dataset was about 820 (bar between $\approx[480,1300]$). The loop was consistent over different choices of the first landmark. We identified 
13 loop members, out of which 9 were found in cancer literature. Some cancer-related members include EGR1 and CAV1, which have genes been characterized as cancer suppressor genes\cite{gene_cav1, gene_egr1}, A2M, which has been identified as a predictor for bone metastases\cite{a2m_2001}, and RPL36 which has been found to be a prognostic marker in hepatocellular carcinoma\cite{rpl36_2011}.

\subsection{Breast dataset}\label{sl:breast}
The lifespan of the longest loop in this dataset is in the range $[10080.0, 16684.2]$. As with the brain dataset, this loop is very consistent. However, there were only 10 members of this loop, and 8 of which were found in cancer literature. An interesting feature of this loop is that it had five ribosomal proteins which are known to play a critical role in tightly coordinating p53 signaling with ribosomal biogenesis \cite{ribo_prots_2011}.

\subsection{Ovarian dataset}\label{sl:ovarian}
The Ovarian dataset had the most variable features in \(\homo_1\). However, we investigated the loop corresponding to the most consistent and longest bar, which ranged from about 4000 to over 7000. This loop consisted of 17 members, and 9 were mentioned in the cancer-related literature. Among cancer-related members were GNAS, which was identified as ``an independent, qualitative, and reproducible biomarker to predict progression-free survival in epithelial ovarian cancer'' \cite{gnas_2010}, and HNRNPA1, which has been identified as a potential biomarker for colorectal cancer \cite{hnrpra1_2009}. 

\subsection{AML188 dataset}\label{sl:aml188}
Acute myeloid leukemia 188 (AML188) had two significant loops (as did AML170). The first one occurred at $[25200.0, 102200.0]$ and the second one at $[78400.0, 146219.24]$. The first loop has 27 members while the second one only 6. Altogether, only 14 of these 33 genes were mentioned in cancer literature. Some cancer-related representatives were hepatic leukemia factor (HLF) which promotes resistance to cell death \cite{hlf_2013}, RPL35A known for inhibition of cell death \cite{rpl35a_2002}, and GRK5 which regulates tumor growth \cite{grk5_2012}. A group of zinc finger proteins were present in the first loop, some of which have been reported as novel biomarkers for detection of squamous cell carcinoma \cite{zink_finger_2011,zink_finger_1991}.

\subsection{AML170 dataset}\label{sl:aml170}
AML170 comes from caArray database and its protocol HG\_U95Av2 has only 12558 genes. Even though this protocol had a smaller (about $1/4$) number of genes compared to the other data sets, we still detected two loops in this dataset. They were relatively shorter than the loops in AML188, and occurred at $[5800.0, 11400.0]$ and $[11000.0, 17600.0]$. The two loops comprised of 19 members, of which 10 were found in cancer-related literature. These relevant members included ubiquitin C (UBC), which was recently identified as a novel cancer-related protein \cite{ubc_2014}, PABPC1 whose positive expression is correlated with tumor progression in esophageal cancer \cite{pabpc1_2006}, and LAMP1 which facilitates lung metastasis \cite{lamp1_2014}.

\section{Discussion}\label{sl:discuss}

\begin{wrapfigure}{r}{3in}
\centering
\includegraphics[width=3in]{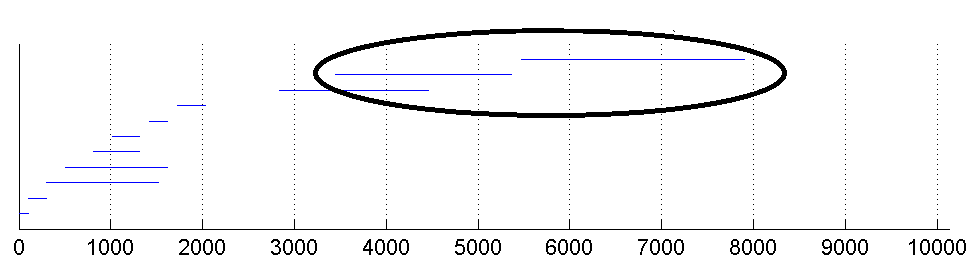}
\caption{Two persistent loops in the AML170 dataset detected using only $25$ dimensions at the number of landmarks L=130.} \label{aml170_dim25}
\end{wrapfigure}
The Breast, Brain, and Ovarian datasets had only one persistent loop, while AML170 and AML188 had two. Also, the AML datasets had a higher number of patients (samples) than the other three sets (see Table~\ref{sl:tbl_data}). Is this fact just a coincidence or, indeed, does the number of \(\homo_1\) features (loops) correlate with the number of dimensions? To address this question, we chose samples of random  and progressively larger ($25$--$175$) subsets of patients from AML170 and AML188 while also increasing the number of landmarks, and studied the evolution of \(\homo_1\) features. In other words, we repeated our method on smaller subsets of patients from these datasets. Both the AML datasets contained two loops even with only 25 dimensions (see Fig.~\ref{aml170_dim25} for AML170), and continued to do so for the progressively larger subsets. Thus, the number of significant \(\homo_1\) features appears to depend on intrinsic qualities of the data rather than the number of dimensions, demonstrating the robustness of our method to the number of patients in the dataset.

An important property of a loop is its lifespan \cite{carlsson2006algebraic}. One may note that the life span of loops for different datasets vary significantly. For example, the lifespan of a significant \(\homo_1\) feature in the brain dataset is only $820$, while for AML188 the lifespan of the first loop is $77,000$. This difference is not only due to the increase in the actual size of a loop as indicated by the number of points comprising the loop ($13$ vs.~$27$ in this case), but also in part because of the different absolute values in microarray expression data. The maximum  value for the brain dataset is \(24 \times 10^3\), while for AML188 is \(3 \times 10^6\). Therefore, the absolute length of an \(\homo_1\) feature is not as important as its length relative to other \(\homo_1\) features.

\begin{figure}[ht!]
\centering
\includegraphics[scale=0.33]{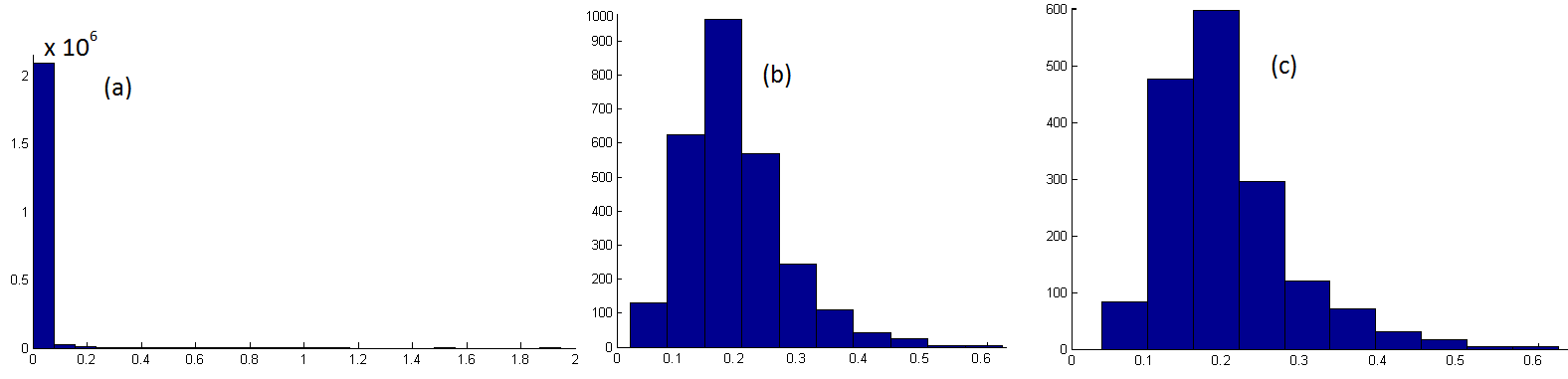}
\caption{Histograms for AML170 dataset, x-axis represents gene expression level of scale $10^4$. (a) distribution of gene expressions for the whole set; (b) distribution of gene expressions for only the loop members; and (c) distribution of gene expressions for cancer-related loop members.} \label{aml170_loop_hist} 
\end{figure}

\begin{figure}[ht!]
\centering
\includegraphics[scale=0.33]{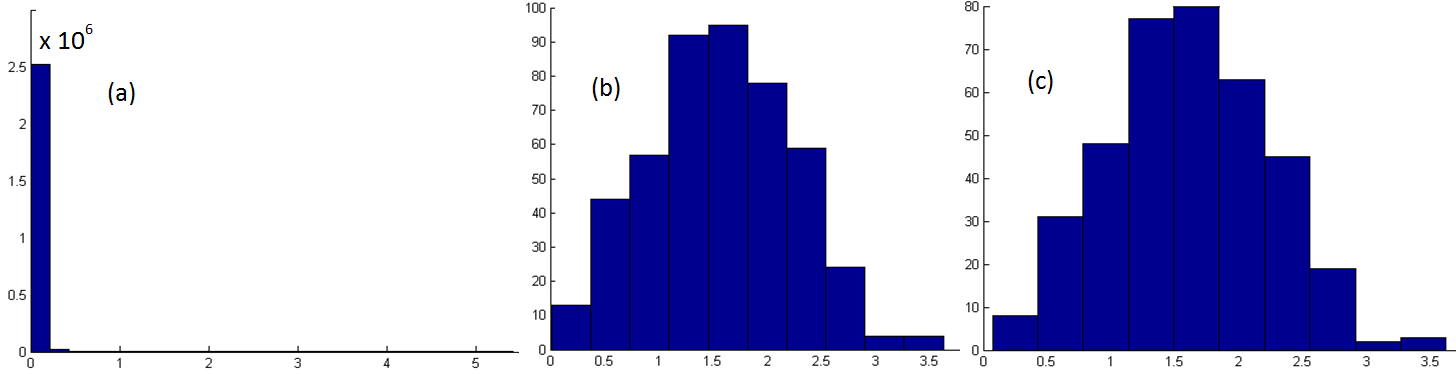}
\caption{\label{breast_loop_hist} Histograms for Breast dataset, x-axis represents gene expression level of scale $10^5$. (a) distribution of gene expressions for the whole set, (b) distribution of gene expressions for the loop members only, (c) distribution of gene expressions for cancer-related loop members.}
\end{figure}

The crucial step of our method is the choice of landmarks. The goal here is the efficient inference of the topology of data, while selecting a small subset of potentially relevant genes. Landmarks chosen using sequential maxmin tend to cover the dataset better and are spread apart from each other, but the procedure is also known to pick outliers \cite{desilva2004}. In our datasets, outliers are typically identified by extreme expression values \cite{outlier_survey2004}. We examined the expression values of the chosen landmarks, and found that very few of them had extreme values (Figs.~\ref{aml170_loop_hist} and \ref{breast_loop_hist}). Similarly, the expressions of the genes implicated in cancer biogenesis (among the loop members) did not have any extreme values, and in fact appear to follow normal distributions. We infer that this observation results because sequential maxmin indeed picks points on the outskirts of the topological features. Further, the distribution of expressions of the loop members suggests that the group as a whole could have potential implications for the disease. More interestingly, the ``hole'' structure means such groups could not potentially be identified by traditional coexpression or even differential expression analyses \cite{ChKe2009}. 
\vspace*{-0.2in}

The computational complexity of our method is based on the current implementation of JavaPlex \cite{javaplex2011}, where the two main steps are building and filtering simplicial complexes, and computing homology. Up to dimension $2$ (\(\beta_2\)), homology could be computed in $O(n^3)$ time \cite{EdHa2009}. However, building simplicial complexes relies on clique enumeration, which is NP-complete, and has a complexity of $O(3^{n/3}) $\cite{jholes2014,cliques65}. Further, JavaPlex requires explicit enumeration of simplicial facets appearing at each filtration step, implying the need for large memory resources\cite{javaplex2011}.

The cancer-related loop members identified from the two AML datasets were distinct apart from the prominent group of ribosomal proteins. This observation could be explained by two main reasons. First, AML188 has four times the number of genes as AML170 (see Table~\ref{sl:tbl_data}). Second, JavaPlex identifies only {\em one} representative of each homology class. That is, if a significant topological feature (hole) exists, it will be identified, but only one loop will be found around that hole. There could be other relevant points in proximity to the hole, but are not guaranteed to be included in the loop. If we have prior knowledge of some genes being relevant, we could try to identify loops around the holes that include these genes as members. In this case, the other members of the identified loops could also have potential implications for the cancer biogenesis. Methods to find a member of a homology class that includes specific points could be of independent interest in the context of optimal homology problems \cite{DeHiKr2011,DeSuWa2010}.

\section{Conclusion}
We have presented a method to look at cancer data from a different angle. Unlike previous methods, we look at characteristics of the first homology group ($\homo_1$). We identify the persistent $\homo_1$ features (which are loops, rather than connected components) and inspect their membership. Importantly, our approach finds potentially interesting connections among genes which cannot be found otherwise using traditional methods. This geometric connectedness may imply functional connectedness, however, this is yet to be investigated by oncologists. If such connections are indeed implied, then the genes in the loops could together form a characteristic ``signature'' for the cancer in question.

\paragraph{Acknowledgment:} 
Krishnamoorthy acknowledges support from the National Science Foundation through grant \#1064600.

\vspace*{-0.1in}
\bibliographystyle{ws-procs11x85}

\end{document}